\documentclass[a4paper,english,reprint,superscriptaddress]{revtex4-2}
\usepackage{lmodern}

\usepackage[T1]{fontenc}
\usepackage{textcomp}
\usepackage[utf8]{inputenc}
\usepackage{parskip}
\usepackage{babel}
\usepackage{float}
\usepackage{booktabs}
\usepackage{amsmath}
\usepackage{amssymb}
\usepackage{graphicx}
\usepackage[pdfusetitle,
 bookmarks=false,
 breaklinks=false,pdfborder={0 0 0},pdfborderstyle={},backref=false,colorlinks=false]
 {hyperref}

\makeatletter

\pdfpageheight\paperheight
\pdfpagewidth\paperwidth

\providecommand{\tabularnewline}{\\}

\makeatother

\begin{document}
\title{Towards experimental demonstration of quantum position verification
using true single photons}
\author{Kirsten Kanneworff}
\email{kanneworff@physics.leidenuniv.nl}

\author{Mio Poortvliet}
\affiliation{Leiden Institute of Physics, Leiden, The Netherlands}
\author{Dirk Bouwmeester}
\affiliation{Leiden Institute of Physics, Leiden, The Netherlands}
\affiliation{Department of Physics, University of California, Santa Barbara, California,
USA}
\author{Rene Allerstorfer}
\affiliation{QuSoft, Amsterdam, The Netherlands}
\affiliation{CWI, Amsterdam, The Netherlands}
\author{Philip Verduyn Lunel}
\affiliation{Sorbonne Université, CNRS, LIP6, France}
\affiliation{QuSoft, Amsterdam, The Netherlands}
\affiliation{CWI, Amsterdam, The Netherlands}
\author{Florian Speelman}
\affiliation{QuSoft, Amsterdam, The Netherlands}
\affiliation{University of Amsterdam, The Netherlands}
\author{Harry Buhrman}
\affiliation{Quantinuum, London, United Kingdom}
\affiliation{QuSoft, Amsterdam, The Netherlands}
\affiliation{University of Amsterdam, The Netherlands}
\author{Petr Steindl}
\author{Wolfgang Löffler}
\email{loeffler@physics.leidenuniv.nl}

\affiliation{Leiden Institute of Physics, Leiden, The Netherlands}
\begin{abstract}
The geographical position can be a good credential for authentication
of a party, this is the basis of position-based cryptography – but
classically this cannot be done securely without physical exchange
of a private key. However, recently, it has been shown that by combining
quantum mechanics with the speed of light limit of special relativity,
this might be possible: quantum position verification. Here we demonstrate
experimentally a protocol that uses two-photon Hong-Ou-Mandel interference
at a beamsplitter, which, in combination with two additional beam
splitters and 4 detectors is rendering the protocol resilient to loss.
With this we are able to show first results towards an experimental
demonstration of quantum position verification.
\end{abstract}
\maketitle
\begin{figure}
\includegraphics[width=1\columnwidth]{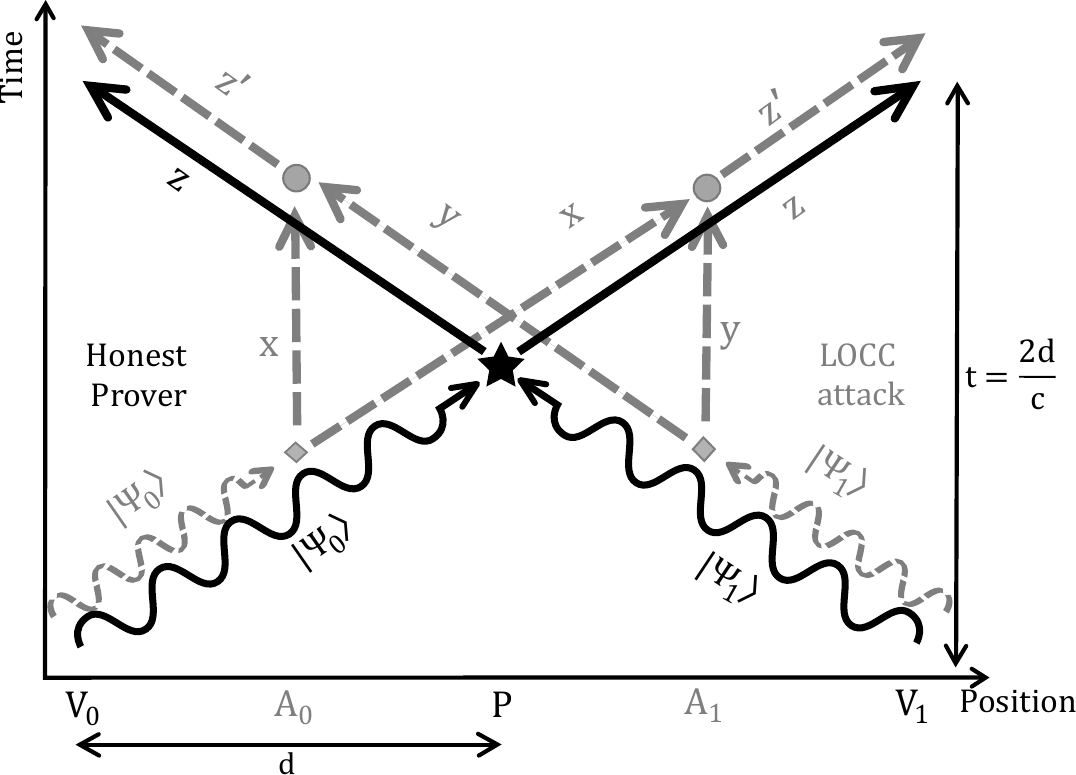}\caption{\label{fig:Space-time_simple}Space-time diagram of a one-dimensional
QPV protocol showing the prover ($P$) centered between two verifiers
($V_{0}$ and $V_{1}$, solid black) where curly (straight) lines
indicate quantum (classical) information exchange. Dashed grey lines
show a potential form of attack by two adversaries ($A_{0}$ and $A_{1}$)
positioned around the supposed location of the prover that try to
mimic the honest prover responses and are restricted to local operations
and classical communication (LOCC). Symbols are explained in the text.}
\end{figure}

Since the geographical location is often a good credential of a party
in communications, verification thereof could add a useful layer to
communication security – this is the case, for instance, with data
centers, banks, government buildings, a lab in a quantum network,
or even a satellite. Classically, position verification is only possible
securely if a shared private key is established which requires physical
contact of the parties \cite{chandran2009}. In quantum mechanics,
mainly thanks to the no-cloning theorem, this can be avoided \cite{kent_tagging_2006,kent_quantum_2011,kent_quantum_2011-1,brassard_conundrum_2011}.
The general scheme of quantum position verification (QPV) is shown
in Fig. \ref{fig:Space-time_simple}: Two verifiers $V_{0}$ and $V_{1}$
share a private communication channel and aim to confirm the location
of a third party, the prover $P$. The verifiers send classical and
quantum information, the prover performs a task and returns classical
(and possibly quantum) information. The verifiers use this information
and the timing and conclude if the prover was at the claimed position
or not. This scheme is one-dimensional but can be extended to higher
dimensions \cite{unruh_quantum_2014}.

However, it quickly was found that attackers with shared entanglement
and exploiting quantum teleportation can break quantum position verification
protocols, after first attempts \cite{kent2006,malaney2010,malaney2010a}
a general attack was found \cite{buhrman2014}. This finding has stimulated
broad research into the topic \cite{qi2015,miller_perfect_2024,amer_certified_2024,escola-farras_quantum_2024,george_orthogonality_2024,may_quantum_2019,olivo_breaking_2020,das_practically_2021,liu_beating_2022,junge_geometry_2022,cree_code-routing_2023,allerstorfer_relating_2024,allerstorfer2022,allerstorfer2022a,lim2016},
and it was found that by including classical-information cryptographic
tasks, QPV protocols can be made secure for all practical purposes
such that attackers require a very large amount of shared entanglement
that does only depend on the amount of classical information used
in the QPV protocol \cite{bluhm2022,asadi_linear_2024}. 

In real-world QPV, the quantum information is sent by photons, and
two major loopholes emerge from this: First, photons are susceptible
to loss during transmission, which opens up a generic attack strategy
since the adversaries can claim loss if their measurements have been
performed in the wrong basis, for instance. Therefore, fully loss-tolerant
protocols are required \cite{allerstorfer2022,escola-farras_single-qubit_2023,escola-farras_lossy-and-constrained_2024},
the first having been developed in Refs. \cite{qi2015,lim2016}. We
will investigate here a variation of those protocols, the SWAP protocol
developed and analyzed by some of us \cite{allerstorfer2022a} where
two-photon interference makes loss-based attacks recognizable. The
second major loophole appears if we transport the photons through
fiber networks, where the speed of light is reduced compared to free
space, giving attackers using free-space communications an advantage.
This we do not address here, but we mention that recently, advanced
protocols including a commitment step have been developed \cite{allerstorfer2023}
that could mitigate this issue in future.

In this paper we report our progress towards an experimental demonstration
of QPV. We use single photons from a demultiplexed quantum dot – microcavity
single-photon source, send them to the two verifiers, encode suitable
qubits in the photons and send them to the prover. The prover performs
the SWAP test using Hong-Ou-Mandel two-photon interference and measures
the result in a loss-tolerant way with 4 single-photon detectors.
We analyze the results critically by comparing photon correlations
to protocol simulations. Those results show that we currently cannot
claim fully-secure QPV, and we find that imperfections in our single-photon
source are responsible that can be avoided in principle as we show.
We conclude with an outlook for future experiments. 

\section{Protocol }

Photon loss is one of the most important limiting factors for any
experimental realization of quantum position verification. Most of
the proposed QPV protocols are partially loss tolerant meaning that
they can only tolerate loss up to a certain fraction such as 50\%.
However, any loss limit renders a real-world implementation very challenging
due to the exponential loss with distance given by the Lambert-Beer
law, and limited photon production and detection probability. The
first ideas about a full loss-tolerant QPV protocol was proposed by
Qi and Siopsis \cite{qi2015} and a first experimental proposal for
such a protocol was developed by Lim et al. \cite{lim2016}. We use
here an adaptation of the latter by Allerstorfer et al. \cite{allerstorfer2022a},
the SWAP protocol, where instead of polarizing beamsplitters non-polarizing
50:50 beamsplitters are used and potentially all 3 mutually unbiased
polarization bases (we show here one basis only).

The SWAP protocol entails, see Fig. \ref{fig:Space-time_simple}:
\begin{enumerate}
\item \textbf{Preparation:} Verifiers $V_{0}$ and $V_{1}$ share via their
private channel a uniformly drawn random sequence of basis choices
and randomly parallel or orthogonal states in the basis, e.g. $|\Psi_{0}\rangle$
and $|\Psi_{1}\rangle$. Encoded in single photons, these qubits are
sent to the prover such that they arrive simultaneously.
\item \textbf{Measurement $\star$: }The prover performs the quantum measurement
based on two-photon Hong-Ou-Mandel (HOM) quantum interference \cite{hong1987}
but with two additional beamsplitters and 4 detectors, which allows
to discriminate HOM photon bunching from loss as explained below.
The prover returns a classical response $z=0$ if $|\Psi_{0}\rangle\parallel|\Psi_{1}\rangle$,
$z=1$ if $|\Psi_{0}\rangle\perp|\Psi_{1}\rangle$, and $z=\oslash$
if the measurement is not conclusive.
\item \textbf{Round check: }After each response of the prover the verifiers
review if the received response $z$ is the same for both verifiers
and if the response arrived within the set time constraint. If either
check fails the verifiers abort the protocol. 
\item \textbf{Verification:} After $n$ rounds of steps $1\dots3$ the verifiers
check if the distributions of answers returned by the prover $z=\{0,1,\o\}$
follows the expected distribution within a certain error margin.
\end{enumerate}

\section{Experiment}

\subsection{The single-photon source}

\begin{figure*}
\includegraphics[width=0.75\textwidth]{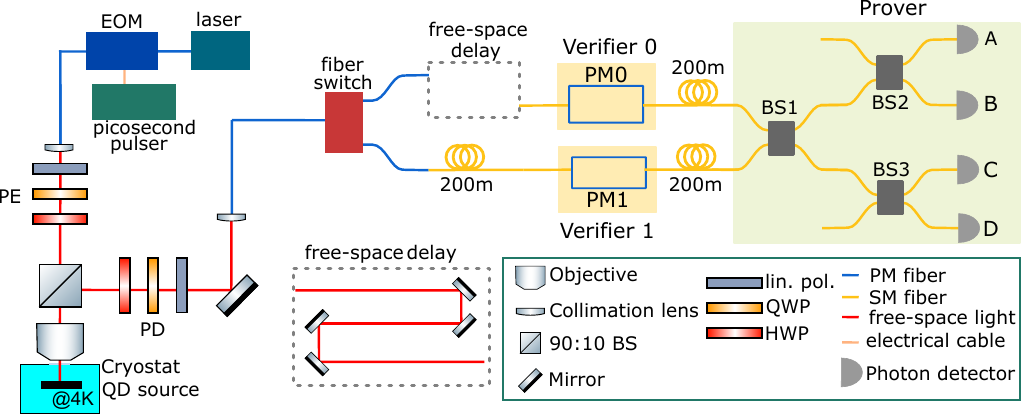}\caption{\label{fig:setup}Schematic of the experimental setup. The electro-optic
modulator (EOM) is used to generate the picosecond pulses, PE and
PD are the polarization control elements of the excitation and detection
paths, PM0 and 1 are the polarization modulators of the verifiers,
and BS1..3 are 50:50 fiber-based beamsplitters. }
\end{figure*}

Essential for our experiment shown in Fig. \ref{fig:setup} is the
source of single photons. We use a single negatively charged self-assembled
InGaAs/GaAs quantum dot (QD) embedded in an optical microcavity \cite{somaschi2016,snijders2018,tomm2021,thomas2021}.
The QD is embedded in a p-i-n junction separated by a 31.8 nm thick
tunnel barrier from the electron reservoir to enable tuning of the
QD resonance wavelength at around 935 nm by the quantum-confined Stark
effect, for details see Refs. \cite{snijders2018,steindl2021,steindl2023a}.
We drive the QD resonantly with short optical pulses carved out of
narrow-linewidth frequency-tunable continuous-wave laser light by
using an electro-optic modulator (EOM) controlled by custom made electronics
\cite{poortvliet2024}. This enables production of laser pulses with
tunable pulse width (of around 100 ps) and pulse period (9 ns) at
a well-defined center wavelength. These parameters provide a good
trade-off between single-photon brightness and quality of the single
photons \cite{poortvliet2024}. The single photons are separated from
the laser light using a cross-polarization technique enabling laser
extinction on the order of $10^{-6}$ \cite{steindl2023} and collected
in a polarization-maintaining (PM) single-mode fiber.

\subsection{QPV setup}

The overall scheme of the QPV experiment is shown in Fig. \ref{fig:setup}.
For the present implementation of the SWAP protocol, we demultiplex
and distribute consecutive single photons from the quantum dot source
to both verifiers. For this, we first temporally demultiplex photons
using a fiber switch (Agiltron NPNS, 1 \textmu s switching time).
The time delay of the demultiplexer setup is adjusted to the switching
frequency, and an additional free-space delay is used to fine-tune
the temporal profile of the single photons to maximally overlap at
the first beamsplitter BS1 of the prover part of the setup. To simulate
the distance between the verifiers and the prover, 200 m of single-mode
optical fiber cable (780HP) is used. The overall transmission of the
setup is between $7.2\%$ and $12.4\%$, details are given in the
supplementary material. We do not implement the classical channel
for returning the prover answers to the verifiers, this can be done
by standard radio-frequency techniques. 

\textbf{Verifiers. }Both verifiers encode their qubits into the polarization
state of the photons using piezo-electric fiber-based polarization
modulators (PM0 and PM1, Polarite III PCD-M02), with which arbitrary
polarization states can be prepared. 

\textbf{Calibration.} All fibers behind the fiber switch are non-polarization
preserving fibers, and all induce polarization rotations. We use a
fiber coupled polarimeter (Thorlabs PAX1000IR1/M) to calibrate the
necessary polarization rotations such that polarization qubits from
both verifiers experience during transmission to the beamsplitter
BS1 the same unitary polarization transformation. To achieve this,
we first replace one detector by the polarimeter, set the switch to
send light through the path of verifier 0, and record the polarization
state. Then we set the switch to direct light through the verifier
1 path, and adjust the polarization modulator PM1 such that the same
polarization state is obtained. In this way we calibrate the transmission
through the full setup and we do not have to change any fiber connections
after this calibration, which avoids unavoidable drifts after reconnecting
or moving a fiber. 

\textbf{Prover.} To realize the SWAP protocol, the prover uses a system
of 3 fiber-based beamsplitters (Thorlabs TW930R5A2) in combination
with four avalanched single-photon detectors (Excelitas SPCM-AQRH-14-FC-ND).
We use a time-tagging card (Cronologic HPTDC, 100 ps resolution) and
custom software to record all single counts and all combinations of
2-, 3-, and 4-fold coincidence detection events. From these coincidence
events, the prover determines their answer, and reports either an
inconclusive result ($z=\oslash$, if zero, one, or more than two
photons are detected) or a conclusive result if two photons are detected.
If conclusive, the prover returns $z=0$ for AB and CD events, i.e.,
if both photons are detected in the same arm after the first beamsplitter
BS1 (and HOM photon bunching probably did happen). If HOM bunching
did not happen, that is for AC, AD, BC and BD events, the prover returns
$z=1$. As mentioned above, adding BS2 and BS3 makes the protocol
resilient against loss since also bunched photons sometimes are detected
as AB or CD coincidence events. 

\section{Results}

\begin{figure}
\includegraphics[width=1\columnwidth]{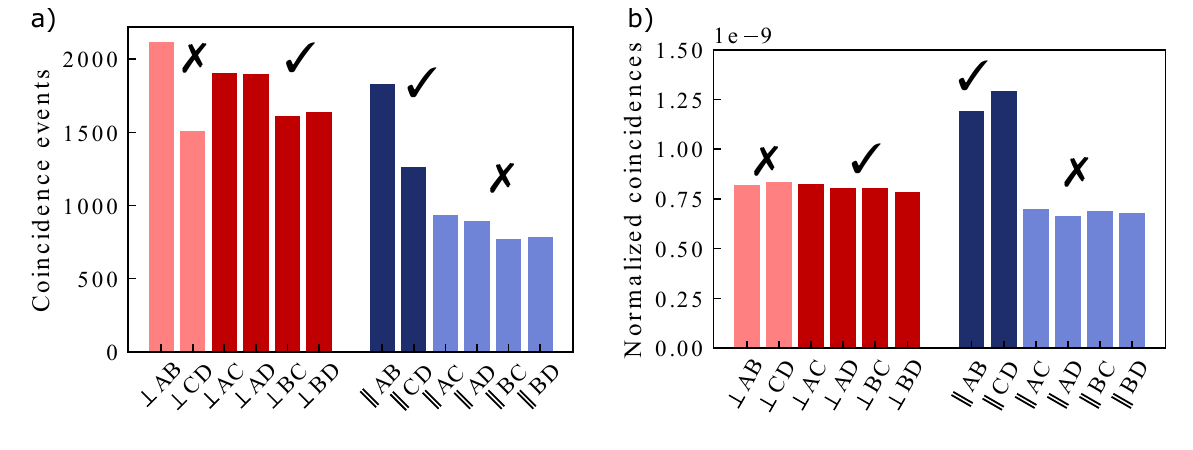}\caption{\label{fig:CCS_SC_raw}Photon correlations at the prover, raw coincidences
$CC_{ij}$ (a) and normalized coincidences $CC_{ij}^{norm}$ (b) for
a 5 hour long measurement. For orthogonal verifier qubits ($\perp$,
red), theory predicts equal rates which is well reproduced in the
experiment. For parallel qubits ($\parallel$, blue), only $\parallel$AB
and $\parallel$CD events are expected - the unwanted events are due
to imperfections of our single-photon source as explained in the text.
}
\end{figure}

The experimental procedure is as follows: (i) We calibrate the polarization
of the setup as described above, and record the settings. (ii) The
single-photon source is optimized (laser power, polarization, quantum
dot bias voltage). (iii) Data is recorded for 5-minute intervals.
Steps (ii) and (iii) are repeated for the measurement time. Fig. \ref{fig:CCS_SC_raw}
shows the raw and normalized coincidence events. We focus here on
only one polarization basis, the HV basis. We note that we observe
no 3- and 4-fold events in our one-hour long measurements.

If our experiment would be perfect, all coincidence events are equally
probable for orthogonal qubits ($\perp$) from the verifiers. This
is well recognizable in Fig. \ref{fig:CCS_SC_raw}, red bars. If the
qubits from the verifiers are equal ($\parallel$), we would expect
perfect HOM photon bunching and that only AB and CD events appear.
In Fig. \ref{fig:CCS_SC_raw}(a), we indeed observe an enhancement
of these events, but also a rather large amount of unexpected coincidences,
which we will discuss below. In Fig. \ref{fig:CCS_SC_raw}(b) we show
the normalized coincidences

\begin{equation}
CC_{ij}^{norm}=\frac{CC_{ij}}{SC_{i}\,SC_{j}}\label{eq:CC_norm}
\end{equation}
where $CC_{ij}$ are the coincidence events of detectors $i$ and
$j$, and $SC_{i}$ are the single photon detection events of detector
$i$. This shows that the large difference between $CC_{AB}^{\parallel}$
and $CC_{CD}^{\parallel}$ in Fig. \ref{fig:CCS_SC_raw}(a) originates
from unbalanced beam splitters and different transmissions of the
respective paths, which is removed by this normalization.

\begin{figure}
\includegraphics[width=1\columnwidth]{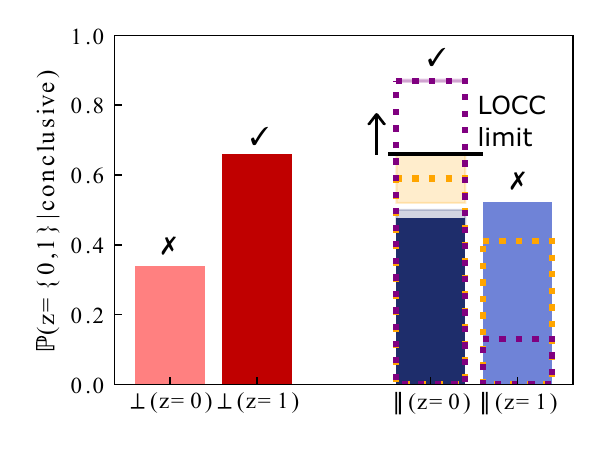}\caption{\label{fig:QPV_HV}Conditional probabilities of prover response for
both orthogonal ($\perp$) and parallel ($\parallel$) qubits sent
from the verifiers, conditioned on the response being conclusive.
The dark colored bars indicate a 'correct' $(\checkmark)$ response
from the prover while the light color indicates an 'incorrect' answer
$(\times)$. The probability of $z=0$ is obtained from the sum of
AB and CD coincidences and the probability for $z=1$ is determined
from the sum of the other four 2-fold coincidences. To obtain probabilities,
both are divided here by the total amount of 2-fold coincidences.
The 'LOCC limit' of $2/3$ is the maximum probability of attackers
responding correctly under the LOCC assumption. Dotted bars show the
modeled conditional probabilities for a better quantum dot single-photon
sources as explained in the text. }
\end{figure}

\textbf{Prover answers.} The prover determines the answer from the
photon detection events as explained above and in the final step in
the verification process the verifiers check if the conclusive responses
from the prover follow the expected distribution. This is done by
calculating the ratio of correct and incorrect answers received from
the prover. We now discuss the expected results, and compare to the
experimental data. The results are shown in Table \ref{tab:g2_QPV_prob}
and Fig. \ref{fig:QPV_HV}.

First, what is the probability to obtain an inconclusive result, where
the two photons are absorbed by the same detector – for the case of
an ideal experiment without loss? In the case of orthogonal qubits
($\perp$) where no HOM photon bunching is happening, the chance that
both photons leave the beamsplitter through the same port is $1/2$,
and this must happen twice, at BS1 and then at BS2 or BS3 - therefore
$\mathbb{P}(\oslash|\perp)$=1/4. In the case of parallel qubits ($\parallel$),
HOM photon bunching happens at BS1 with certainty, and therefore the
chance of an inconclusive result is twice as high: $\mathbb{P}(\oslash|\parallel)=1/2$.

Now, we discuss the different probabilities conditioned on a conclusive
answer, i.e., that two photons were detected. For the case of orthogonal
qubits ($\perp$) arriving from the verifiers, since no HOM photon
bunching happens, all 6 coincidence events are equally probable. We
obtain $\mathbb{P}(0|\perp,\text{concl.})=2/6=1/3$ and $\mathbb{P}(1|\perp,\text{concl.})=4/6=2/3$.
This is important, also in the ideal case, the prover will return
the ``wrong'' answer $z=0$ that should indicate parallel qubits.
Finally, for parallel $\parallel$ qubits, the photons exit BS1 through
the same port as a consequence of HOM photon bunching, only AB and
CD coincidences can occur which results in $z=0$ and consequently
$\mathbb{P}(0|\parallel,\text{concl.})=1$.

\begin{table}
\begin{tabular*}{1\columnwidth}{@{\extracolsep{\fill}}ccc}
\toprule 
 & Theory & Experiment\tabularnewline
\midrule
\midrule 
$\mathbb{P}(\oslash|\perp)$ & $1/4$ & NA\tabularnewline
\midrule 
$\mathbb{P}(\oslash|\parallel)$ & $1/2$ & NA\tabularnewline
\midrule 
$\mathbb{P}(0|\perp,\text{concl.})$ & $1/3$ & 0.34\tabularnewline
\midrule 
$\mathbb{P}(1|\perp,\text{concl.})$ & $2/3$ & 0.66\tabularnewline
\midrule 
$\mathbb{P}(0|\parallel,\text{concl.})$ & 1 & 0.48\tabularnewline
\midrule 
$\mathbb{P}(1|\parallel,\text{concl.})$ & 0 & 0.52\tabularnewline
\bottomrule
\end{tabular*}\caption{\label{tab:g2_QPV_prob}Expected and measured probabilities.}
\end{table}
These expectations and the experimental results calculated from the
data in Fig. \ref{fig:CCS_SC_raw} are shown in Table \ref{tab:g2_QPV_prob}
and Fig. \ref{fig:QPV_HV}. 

\textbf{LOCC attack.} We now sketch which best-case probabilities
two adversaries can obtain, if they are restricted to LOCC. Every
round, each adversary intercepts (see Fig. \ref{fig:Space-time_simple})
the qubit sent by the verifier closest to them and measures it in
a certain basis (diamonds in Fig. \ref{fig:Space-time_simple}). Then,
they share their results with the other adversary and formulate a
response that is sent to the verifiers (circles in Fig. \ref{fig:Space-time_simple}).
Assuming that the verifiers use all three mutually unbiased bases,
there is a $1/3$ probability that the adversaries have measured in
the correct basis which enables them to return the correct expected
result with certainty. For the other two basis choices (each also
occurring with a $1/3$ probability), there is still a chance of $1/2$
to guess correctly the answer, therefore we obtain as the correct-guessing
probability of the LOCC adversaries 
\begin{equation}
\mathbb{P}_{\text{succes}}^{\text{LOCC}}=\frac{1}{3}\left(1+\frac{1}{2}+\frac{1}{2}\right)=\frac{2}{3}.\label{eq:prob_LOCC}
\end{equation}
A proper proof for this bound is given in Ref. \cite{allerstorfer2022a}.
As mentioned before, even in an ideal experiment and without adversaries,
for orthogonal qubits, the result is correct with only a chance of
$2/3$. Since, however, ideally, equal amounts of rounds are played
with orthogonal and parallel qubits, where the latter results always
in the correct answer, the correct answer is sent with probability
$5/6$.

\section{Discussion}

For orthogonal qubits $(\perp)$ the measurement data follows the
expected distribution where $2/3$ of the time the honest prover responds
correctly as seen in Fig. \ref{fig:QPV_HV}, and we conclude that
differences in efficiencies in the setup they are not significant
for the prover responses in this case. For parallel $(\parallel)$
qubits, as we have mentioned, our data deviates from the expectations,
the origin of this we explore now.

We have made a simple model of our experiment including photon source
parameters, and all characteristics of the optical setup including
loss, unbalanced fiber beam splitters, and detection efficiencies,
a detailed characterization is given in the supplemental information.
The single-photon source is characterized by the the single-photon
purity $P$ and the photon indistinguishability or wave-function overlap
$M$ \cite{somaschi2016,tomm2021,thomas2021,ding_high-efficiency_2023}
- we ignore the single-photon brightness here. The single-photon purity
$P$ is given by $P=1-g^{(2)}$ where the zero-time second-order correlation
function $g^{(2)}$ is measured in a Hanbury-Brown and Twiss type
experiment. To obtain the wavefunction overlap $M$, we first measure
in a Hong-Ou-Mandel type experiment the zero-time second-order correlation
functions for orthogonal ($g_{\perp,HOM}^{(2)}$) and parallel and
($g_{\parallel,HOM}^{(2)}$) polarized photons. From this, the interferometric
Hong-Ou-Mandel visibility $\mathcal{V}_{HOM}$ can be obtained from
\cite{patel2008} 
\begin{equation}
\mathcal{V}_{HOM}=\frac{g_{\perp}^{(2)}-g_{\parallel}^{(2)}}{g_{\perp}^{(2)}}.\label{eq:HOMvis}
\end{equation}
Now we can calculate the bare photon indistinguishability or wave-function
overlap from \cite{tomm2021}
\begin{equation}
M=\mathcal{V}_{HOM}\left(1+2g^{(2)}\right),\label{eq:indis}
\end{equation}
which shows that the interferometric visibility $\mathcal{V}_{HOM}$
is reduced by a non-ideal single-photon purity. 

For our source, we measure $g_{\parallel,HOM}^{(2)}=(36.8\pm3.0)\%$
and $g_{\perp,HOM}^{(2)}=(58.8\pm3.6)\%$, resulting in a interferometric
visibility of $\mathcal{V}_{HOM}=(37.4\pm6.4)\%$ and an indistinguishability
of $M=(54.2\pm10.1)\%$. To figure out the origin of our non-ideal
result above, and to identify where our experiment can most easily
be improved, we use our model to predict the most critical QPV probability
$\mathbb{P}(0|\parallel,\text{concl.})$, i.e. that the prover answers
$z=0$ on parallel inputs $|\Psi_{0}\rangle\parallel|\Psi_{1}\rangle$.
We use all our experimental details but alter the single photon performance
metrics - using experimental data from an excellent single photon
source by Tomm et al. \cite{tomm2021}. We consider two cases in addition
to ours (A), first using all metrics from Tomm et al. (B), and then
only their single photon purity but our indistinguishability (C).
In each case, indistinguishability data of photons produced 1 \textmu s
apart are used. All results are shown in Table \ref{tab:g2_QPV}.
We see that a near-ideal single-photon source (case B, also indicated
by the purple bar in Fig. \ref{fig:QPV_HV}) is sufficient to clearly
exceed the threshold of $\mathbb{P}(0|\parallel,\text{concl.})=2/3$,
but also just an improved purity would bring our experiment closer
to this threshold (case C, orange bar in Fig. \ref{fig:QPV_HV}).
In our case, this is caused by non-resonant background emission, finite
cross-polarizastion laser extinction, and by re-excitation of the
quantum dot since the length of the excitation pulse was similar to
the QD lifetime.

\begin{figure}
\includegraphics[width=1\columnwidth]{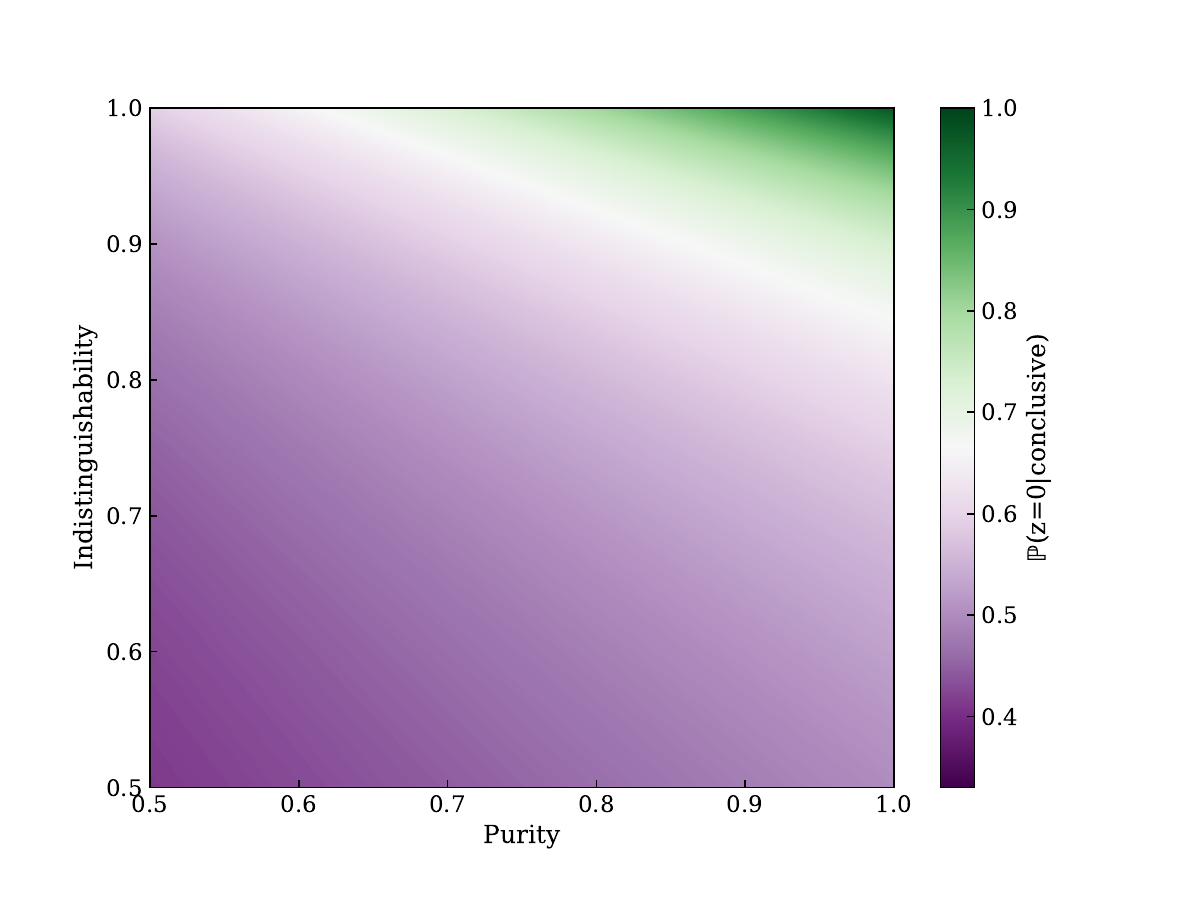}\caption{Probability of a correct $z=0$ response $\mathbb{P}(0|\parallel,\text{concl.})$
depending on single-photon purity and indistinguishability using otherwise
our experimental parameters. The white line marks the threshold of
$2/3$ above which (green) a LOCC attack is not successful.\label{fig:pur-ind-qpv}}
\end{figure}

\begin{table}
\begin{tabular*}{1\columnwidth}{@{\extracolsep{\fill}}cccc}
\toprule 
 & Here (A) & Tomm et al. (B) & Mix (C)\tabularnewline
\midrule
\midrule 
Purity $P$ & $0.776\pm0.017$ & $0.979\pm0.001$ & $0.979\pm0.001$\tabularnewline
\midrule 
$g_{\parallel}^{(2)}(0)$ & $0.368\pm0.030$ &  & \tabularnewline
\midrule 
$g_{\perp}^{(2)}(0)$ & $0.588\pm0.036$ &  & \tabularnewline
\midrule 
$\mathcal{V}_{HOM}$ & $0.374\pm0.064$ & $0.916\pm0.001$ & $0.520\pm0.100$\tabularnewline
\midrule 
$M$ & $0.542\pm0.101$ & $0.960\pm0.005$ & $0.542\pm0.101$\tabularnewline
\midrule 
$\mathbb{P}(0|\parallel,\text{concl.})$ & $0.47\pm0.03$ & $0.870\pm0.003$ & $0.59\pm0.07$\tabularnewline
\bottomrule
\end{tabular*}\caption{\label{tab:g2_QPV}Overview of the parameters and resulting conditional
probability $\mathbb{P}(0|\parallel,\text{concl.})$ for our single-photon
source (A), the source presented in Tomm et. al. (B, \cite{tomm2021})
and for a source similar to our (A) but with improved single-photon
purity (C). }
\end{table}
Finally, we show in Fig. \ref{fig:pur-ind-qpv} how the probability
$\mathbb{P}(0|\parallel,\text{concl.})$ depends on the single-photon
purity and indistinguishability, where otherwise our experimental
parameters and inaccuracies given in the Supplemental Material Section
A are used. We see that both purity and indistinguishability need
to be high to exceed the threshold of $2/3$.

\section{Conclusions and Outlook}

We have shown first experimental results for a loss-tolerant quantum
position verification protocol, using a temporally demultiplexed quantum
dot - microcavity based single-photon source. We found that the Hong-Ou-Mandel
visibility of our single-photon source is the limiting factor to reach
the threshold for quantum secure discrimination between a honest prover
and adversaries that are restricted to local operations and classical
communication (LOCC), i.e., not having shared entanglement. We also
found that with an improved single photon source, this threshold can
easily be reached. For future research, next to improvements of the
single photon source, we stress that addressing the slow quantum information
loophole is most urgent as it would allow using existing fiber networks,
and a promising candidate is a functional single-photon QPV protocol
\cite{bluhm2022} in combination with a commitment step \cite{allerstorfer2023}.

\section*{Acknowledgements}

We acknowledge funding from NWO/OCW (Quantum Software Consortium,
Nos. 024.003.037, 024.003.037/3368), from the Dutch Ministry of Economic
Affairs (Quantum Delta NL, No. NGF.1582.22.025), and from the European
Union’s Horizon 2020 research and innovation program under Grant Agreement
No. 862035 (QLUSTER).

\bibliographystyle{naturemagwV1allauthors}
\bibliography{QPV_manuscript_3,QPV_manuscript_3_WL}

\clearpage \onecolumngrid\renewcommand{\thefigure}{S\arabic{figure}}\setcounter{figure}{0}\renewcommand{\theequation}{S\arabic{equation}}\setcounter{equation}{0}\renewcommand{\thetable}{S\arabic{table}}\setcounter{table}{0}

\section*{Supplemental Information}

\subsection{Experimental setup characterization}

\begin{figure}[H]
\centering{}\includegraphics[width=0.65\paperwidth]{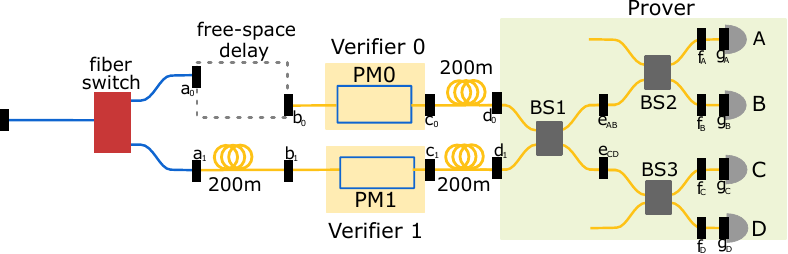}\caption{\label{fig:setup_throughput}Sketch of a part of the experimental
setup with labels indicating the measurement points for the characterization.}
\end{figure}

Here we present a precise characterization of the experimental setup,
which is crucial for the model used in the main text. For this, we
directly connected a continuous-wave (CW) laser to the input of the
fiber switch using the wavelength of the single photons (around 935
nm) and measure the intensity of the laser light at every fiber connection
with a power meter (Thorlabs PM100D). The position of every fiber
connection is depicted in Fig. \ref{fig:setup_throughput} and the
measured transmission ratios are shown in Table \ref{tab:Overview-of-transmission}.
For intensity measurements behind BS1, we blocked the beam in the
free-space delay stage to avoid interference effects. The transmission
ratios given for the beamsplitters (BS1, BS2 and BS3) are the ratios
between the input intensity and the sum of the intensities at the
two outputs of the beamsplitter. The splitting ratios are presented
in Table \ref{tab:beamsplitter_ratio}. The overall efficiency of
the system is between $7.2\%$ and $12.4\%$ and depends on the path
taken and detection efficiency of the detectors.

\begin{table}[H]
\centering{}%
\begin{tabular}{ccc}
\toprule 
 & Transmission (\%) & Transmission (\%)\tabularnewline
\midrule
\midrule 
 & Verifier 0  & Verifier 1\tabularnewline
\midrule 
after switch (a) & 71.2 & 60.3\tabularnewline
\midrule 
delay stage (b) & 95.4 & 91.5\tabularnewline
\midrule 
polarization modulator (PM) (c) & 81.4 & 89.4\tabularnewline
\midrule 
200m fiber transmission  (d) & 86.2 & 85.2\tabularnewline
\midrule 
total (a-d) & 47.7 & 42.0\tabularnewline
\midrule 
BS1{*} (e) & 94.9 & \tabularnewline
\midrule 
BS2{*} ($\text{f}_{\text{A/B}}$) & 99.7 & \tabularnewline
\midrule 
BS3{*} ($\text{f}_{\text{C/D}}$) & 86.8 & \tabularnewline
\midrule 
detector A fiber ($\text{g}_{\text{A}})$ & 90.6 & \tabularnewline
\midrule 
detector A efficiency{*}{*} & 100 & \tabularnewline
\midrule 
detector B fiber ($\text{g}_{\text{B}})$ & 90.3 & \tabularnewline
\midrule 
detector B efficiency{*}{*} & 61.9 & \tabularnewline
\midrule 
detector C fiber ($\text{g}_{\text{C}})$ & 90.7 & \tabularnewline
\midrule 
detector C efficiency{*}{*} & 68.9 & \tabularnewline
\midrule 
detector D fiber ($\text{g}_{\text{D}})$ & 97.9 & \tabularnewline
\midrule 
detector D efficiency{*}{*} & 15.9 & \tabularnewline
\bottomrule
\end{tabular}\caption{\label{tab:Overview-of-transmission}Overview of relative transmissions
for each component in the experimental setup as shown in Fig. \ref{fig:setup_throughput}.
The loss of the fiber-based beamsplitters ({*}) is measured as the
ratio between the input of the beamsplitter and the sum of the two
outputs. The splitting ratios themselves are described in Table \ref{tab:beamsplitter_ratio}.
All detector efficiencies ({*}{*}) are normalized to that of detector
A.}
\end{table}
\begin{table}[H]
\centering{}%
\begin{tabular}{ccc}
\toprule 
Beamsplitter & Ratio upper output (\%) & Ratio lower output (\%)\tabularnewline
\midrule
\midrule 
BS1 $(\text{d}_{1})$ & 54.5 $(\text{e}_{\text{AB}})$ & 45.5 $(\text{e}_{\text{CD}})$\tabularnewline
\midrule 
BS2 $(\text{e}_{\text{AB}})$ & 44.1 $(\text{f}_{\text{A}})$ & 55.9 $(\text{f}_{\text{B}})$\tabularnewline
\midrule 
BS3 $(\text{e}_{\text{CD}})$ & 53.0 $(\text{f}_{\text{C}})$ & 47.0 $(\text{f}_{\text{D}})$\tabularnewline
\bottomrule
\end{tabular}\caption{\label{tab:beamsplitter_ratio}Overview of the splitting ratios of
the fiber-based beamsplitters (Thorlabs TW930R5A2), not accounting
for the total loss in transmission described in Table \ref{tab:Overview-of-transmission}.
The labels in brackets denotes between which points in the setup the
ratios were measured.}
\end{table}

\subsection{Measured coincidence events and normalized coincidences}

\begin{table}[H]
\centering{}%
\begin{tabular}{ccccc}
\toprule 
 & \multicolumn{2}{c}{Coincidence events} & \multicolumn{2}{c}{Normalized coincidences}\tabularnewline
\midrule
\midrule 
 & $\perp$ & $\parallel$ & $\perp/10^{-9}$ & $\parallel/10^{-9}$\tabularnewline
\midrule 
AB & 2115 & 1833 & 0.82 & 1.19\tabularnewline
\midrule 
CD & 1512 & 1261 & 0.83 & 1.29\tabularnewline
\midrule 
AC & 1906 & 934 & 0.82 & 0.70\tabularnewline
\midrule 
AD & 1897 & 893 & 0.81 & 0.67\tabularnewline
\midrule 
BC & 1610 & 770 & 0.81 & 0.69\tabularnewline
\midrule 
BD & 1640 & 784 & 0.79 & 0.68\tabularnewline
\bottomrule
\end{tabular}\caption{\label{tab:coincidences} Overview of values reported in Fig. \ref{fig:CCS_SC_raw}.}
\end{table}

\end{document}